\documentclass[article,prl,twocolumn,nofootinbib]{revtex4}

\usepackage{ifpdf}
\usepackage[T1]{fontenc}
\usepackage[latin9]{inputenc}
\usepackage{amsmath}
\usepackage{amssymb}
\usepackage{graphicx}
\usepackage{subfig}
\usepackage{setspace} 
\usepackage{comment}
\usepackage{tabularx} 

\makeatletter

\@ifundefined{textcolor}{}
{%
 \definecolor{BLACK}{gray}{0}
 \definecolor{WHITE}{gray}{1}
 \definecolor{RED}{rgb}{1,0,0}
 \definecolor{GREEN}{rgb}{0,1,0}
 \definecolor{BLUE}{rgb}{0,0,1}
 \definecolor{CYAN}{cmyk}{1,0,0,0}
 \definecolor{MAGENTA}{cmyk}{0,1,0,0}
 \definecolor{YELLOW}{cmyk}{0,0,1,0}
 \definecolor{PURPLE}{rgb}{0.5,0,0.5}
 }

\newcommand{\bite}{\begin{itemize}}
\newcommand{\eat}{\end{itemize}}
\newcommand{\beq}{\begin{equation}}
\newcommand{\eeq}{\end{equation}}

\newcommand{\beqa}{\begin{align}}
\newcommand{\eeqa}{\end{align}}
\newcommand{\barr}{\begin{array}}
\newcommand{\earr}{\end{array}}

\newcommand{\mb}[1]{\mathbf{#1}}
\newcommand{\mc}[1]{\mathcal{#1}}
\newcommand{\mbb}[1]{\mathbb{#1}}
\newcommand{\mf}[1]{\mathfrak{#1}}

\newcommand{\udarrow}{{\uparrow\downarrow}}

\newcommand{\vect}[1]{\boldsymbol{#1}}

\newcommand{\subsc}[1]{$_{\textrm{#1}}$}

\newcolumntype{V}{>{\centering\arraybackslash} m{.4\linewidth} }

\makeatother

\begin{document}


\title{Superconducting and Anti-Ferromagnetic Phases of Spacetime}

\author{Deepak Vaid}
\email{dvaid79@gmail.com}
\affiliation{National Institute of Technology, Karnataka (NITK)}

\date{\today}

\begin{abstract}
A correspondence between the $SO(5)$ theory of High-T\subsc{C} superconductivity and antiferromagnetism, put forward by Zhang and collaborators, and a theory of gravity arising from symmetry breaking of a $SO(5)$ gauge field is presented. A physical correspondence between the order parameters of the unified SC/AF theory and the generators of the gravitational gauge connection is conjectured. A preliminary identification of regions of geometry, in solutions of Einstein's equations describing charged-rotating black holes embedded in deSitter spacetime, with SC and AF phases is carried out.
\end{abstract}

\maketitle


%

\section{Introduction}

Two of the outstanding problems in theoretical physics today are those of high temperature superconductivity (HTC) on the one hand and quantum gravity (QG) on the other. In the case of HTC, it has been demonstrated that the anti-ferromagnetic (AF) and superconducting (SC) phases $ d $-wave \cite{Demler1995Theory,Demler2004SO5-theory,Zhang1996SO5-Quantum,Zhang1997Unified} and $ p $-wave \cite{Murakami1999An-SO5-model} superconductors can be given a unified explanation in terms of a non-linear sigma model for a field which behaves as a vector transformation under $ SO(5) $ rotations. In QG research it is known that general relativity with non-zero cosmological constant ($ \Lambda \ne 0 $)can be obtained from a so-called BF model (a topological field theory) for a gauge field, valued in either $ SO(3,2) $ (for $ \Lambda < 0 $)  or a $ SO(4,1) $ (for $ \Lambda > 0 $), by a symmetry breaking mechanism \cite{Freidel2005Quantum,Wise2009MacDowellMansouri}. This mechanism was first outlined in a seminal paper by MacDowell and Mansouri \cite{MacDowell1977Unified} in 1977 whose motivation was to construct a unified theory of gravity and supergravity. Similar work was undertaken by Stelle and West in 1980 \cite{Stelle1980Spontaneously}.

More recent work on this topics are the papers by Freidel and Starodubstev \cite{Freidel2005Quantum} Randono \cite{Randono2010Gauge,Randono2010Gravity}, and Westman, Zlosnik and collaborators \cite{Magueijo2013Cosmological,Westman2013Gravity,Westman2015An-introduction}. The notion that geometry should have various phases is suggested by the numerical work in the field of Causal Dynamical Triangulations (CDT) \cite{Ambjorn2013Quantum}. Moreover in \cite{Ansari2008A-statistical,Ansari2008The-Statistical} Mohammad Ansari has demonstrated a connection between anti-ferromagnetism and a statistical formulation of CDT. The connection between the spin-networks used in Loop Quantum Gravity and the Ising model has recently been discussed in \cite{Dittrich2013Ising} and \cite{Feller2015Ising}.

What is new in the present work, to the best of our knowledge, is that it is the first to connect the symmetry breaking on the gravitational side with a well-established model on the condensed matter side. In any situation where a symmetry is spontaneously broken, it is crucial to not only be able to identify the underlying microscopic dynamics which causes the symmetry to break and also to be able to identify and classify the various phases that result from this process. Here we are able to take the first tentative steps towards achieving both these goals.

In this work we demonstrate the equivalence between these two theoretical frameworks. The picture resulting from our line of reasoning is that of a spacetime with non-zero $ \Lambda $, as described by classical general relativity, emerging via symmetry breaking of a topological quantum field theory (TQFT) defined on a four-dimensional manifold. The superconducting phase can be identified with the near horizon geometry of a charged black hole and the antiferromagnetic phase can be identified with the geometry far from the horizon of a rotating black hole. A mixed phase, consisting of both SC and AF phases, would correspond to the spacetime of a charged, rotating black hole.

\section{$ SO(5) $ model of High Temperature Superconductivity}\label{sec:so5_model}

One can ask why the group $SO(5)$ should have anything to do with the description of the SC or AF phases in condensed matter systems, and for that matter, why do we need a unified description of the two phases in the first place. There are three reasons \cite{Zhang1997Unified} to believe that this should be the case:
\begin{enumerate}
	\item In 1988, Chakravarty, Halperin and Nelson \cite{Chakravarty1989Two-dimensional} demonstrated that the non-linear sigma model for a field with $SO(3)$ symmetry gives a good description of the properties of a two-dimensional (2+1) Heisenberg anti-ferromagnet in the low-temperature, long wavelength regime.
	\item The behavior of the superconducting state is known to be well-described by a so-called ``XY'' model for a $U(1)$ gauge field.
	\item Both d-wave SC and AF can be described in terms of the behavior of singlet pairs in the Hubbard model at half-filling. These singlet pairs can describe either an AF phase, a SC phase or a so-called ``spin-bag'' phase where both the phases co-exist.
\end{enumerate}

Now, if both AF and SC arise in different regimes of a system with the same underlying physics - the Hubbard model at half-filling - and can co-exist under certain conditions, it follows, that one would be well-advised to seek out a low-temperature, long wavelength effective field theory which can describe both phases. Such a theory should contain a $SO(3) \times U(1)$ symmetry, which should arise after some symmetry breaking transition. The smallest gauge group that can accomodate such a symmetry among its subgroups is $SO(5)$.

In \cite{Demler1995Theory,Demler2004SO5-theory,Murakami1999An-SO5-model,Zhang1996SO5-Quantum,Zhang1997Unified} it was shown that a non-linear sigma model for a field with an $SO(5)$ gauge symmetry, can describe the physics of both the AF and SC phases. Four of the elements of the generators of the group algebra can be identified with the AF and SC order parameters. The remaining six can be identified with operators which generate transformations between the AF and SC phases.

As argued by Zhang \cite{Zhang1997Unified}, the physical picture of the transition between the AF and SC states is the following. The overall system is described by some microscopic Hamiltonian describing the interaction between electrons on a lattice. Below some characteristic temperature $T_{MF}$, electrons on neighboring sites tend to from singlet bound pairs or \emph{dimers}. The AF and the SC phase are different states which this dimer collective can form. When the dimers are not free to move - due to the lack of vacancies on the lattice - the collective forms a dimer ``solid'' which corresponds to the AF phase. As a certain system parameter is varied the dimer solid begins to melt and forms a fluid which corresponds to the SC phase. At the transition between the two phase one will have regions where both the solid and liquid phases are present. This corresponds to the ``spin-bag'' phase where both AF and SC co-exist.

Let us introduce some notation. $ c^\dagger_{\vect{p},\udarrow} $ ($ c_{\vect{p}, \udarrow} $)is the operator which creates (resp. destroys) an electron with momentum $ \vect{p} $ and given spin ($\uparrow$ or $\downarrow$). $ \sigma^i, \{i=x,y,z\} $ are the Pauli spin-matrices. With these in hand we can define the operators for spin ($ S $), momenta ($ \pi $) and total charge ($ Q $) for electrons near the Fermi surface. These are as follows:
\begin{subequations}
\label{eqn:so5-operators}
	\begin{align}
		S_i & = \sum_{\vect{k},\alpha,\beta} c^\dagger_{\vect{k}\alpha} \sigma^i_{\alpha\beta} c_{\vect{k},\beta} \\
		\pi_i(\vect{k}) & = \frac{1}{4}\sum_{\vect{p},\alpha,\beta} g(\vect{k},\vect{p}) c_{-\vect{p},\alpha}(\sigma^y\sigma^i)_{\alpha\beta} c_{\vect{p},\beta} \\
		Q & = \frac{1}{2} \sum_{\vect{k},\alpha} \left( c^\dagger_{\vect{k},\alpha} c_{\vect{k},\alpha} - \frac{1}{2} \right)
	\end{align}
\end{subequations}
Here $ g(\vect{k},\vect{p})\equiv g(\vect{k}-\vect{p}) $ is a function of the electron momenta in terms of which the operator for the superconducting gap $ \Delta $ can be written as \cite{Zhang1996SO5-Quantum,Zhang1997Unified}:
\begin{equation}
	\label{eqn:gap-operator}
	\Delta^\dagger = \frac{1}{2}\sum_{\vect{k}} g(\vect{k}) c^\dagger_{\vect{k}\uparrow} c^\dagger_{-\vect{k}\downarrow}
\end{equation}
Thus $ g(\vect{k}) $ possesses the symmetries of the gap function $ \Delta_{\vect{k}} $. For the case of d-wave HTSC, it has the form:
\begin{equation}
	g(\vect{k}) = \cos k_x - \cos k_y
\end{equation}
These operators can be arranged in the form of a $ 5 \times 5 $ matrix $ L_{IJ} $ as in:
\begin{equation}\label{eqn:so5-generators-1}
		L_{IJ}	=	\left( \begin{array}{ccccc}
							 0 &  &  &  & \\ 
							 \pi^\dagger_x + \pi_x & 0 &  &  & \\ 
							 \pi^\dagger_y + \pi_y & -S_z & 0 &  &  \\ 
							 \pi^\dagger_z + \pi_z & S_y & -S_x & 0 &  \\ 
				 			 Q & -i(\pi^\dagger_x - \pi_x) & -i(\pi^\dagger_y - \pi_y) & -i(\pi^\dagger_z - \pi_z) & 0
						\end{array} \right) 
\end{equation}
where $ L_{IJ} = -L_{JI} $. It can be shown that, if $ |g(\vect{k})|^2 = 1 $, the elements of this matrix satisfy the commutation relations:
\begin{equation}\label{eqn:so5-commutation}
	\left[ L_{IJ}, L_{KL} \right] = i \left(\delta_{IK} L_{JL} - \delta_{IL} L_{JK} - \delta_{JK} L_{IK} + \delta_{IL} L_{JK} \right)
\end{equation}
which are the commutation rules satisfied by the generators of the Lie alebgra of the group $ SO(5) $. Furthermore the vector $ \vect{S} = (S_x,S_y,S_z) $ can be identified as the order parameter of the ferro/anti-ferromagnetic (F/AF) phase, while $ Q $ is the order parameter of the superconducting (SC) phase. The operator $ \vect{\pi} = (\pi_x, \pi_y, \pi_z) $ rotates the AF order parameter into the SC order parameter and vice-versa \cite{Zhang1997Unified}.

Zhang \cite{Zhang1996SO5-Quantum,Zhang1997Unified} suggests that the behavior of high-T\subsc{TC} superconductors can be characterized by introducing a five-dimensional superspin vector $n^I$, whose components can be identified with the various AF and SC order parameters as follows:

\beq
{\setstretch{2}
\label{eqn:superspin}
	\begin{array}{ccc}	
		n^1 & = & \Delta^\dagger + \Delta \\
	
		n^5 & = & -i(\Delta^\dagger - \Delta) \\
	
		n^2 & = & \displaystyle \sum_{\vect{k},\alpha,\beta} c^\dagger_{\vect{k} + \vect{q},\alpha} \sigma^x_{\alpha\beta} c_{\vect{k},\beta}\\
	
		n^3 & = & \displaystyle \sum_{\vect{k},\alpha,\beta} c^\dagger_{\vect{k} + \vect{q},\alpha} \sigma^y_{\alpha\beta} c_{\vect{k},\beta}\\
	
		n^4 & = & \displaystyle  \sum_{\vect{k},\alpha,\beta} c^\dagger_{\vect{k} + \vect{q},\alpha} \sigma^z_{\alpha\beta} c_{\vect{k},\beta}\\
	\end{array}	
}
\eeq
where $\vect{q} = (\pi_x,\pi_y,\pi_z)$ is the AF order parameter; the operators $\Delta, \pi$ have been defined previously in \eqref{eqn:so5-operators} and \eqref{eqn:gap-operator}.

The matrix $L_{IJ}$ and the vector $n^I$ satisfy the following commutation relations:
\begin{equation}\label{eqn:so5-commutation-b}
	[L_{IJ},n_K] = -i (\delta_{IK} n_J - \delta_{JK} n_I)
\end{equation}
and can be seen to be conjugate variables \cite{Demler2004SO5-theory}, just as the momentum and position $p,q$ are in the ordinary harmonic oscillator. Thus in terms of these objects we can write down the Hamiltonian for $SO(5)$ effective theory of AF and SC:
\begin{align}\label{eqn:so5-hamiltonian}
	H_{SO5} & = \frac{1}{2\chi} \sum_{x,I<J} L^2_{IJ}(x) + \frac{\rho}{2} \sum_{<x,x'>,a} n^I(x) n_I(x') \nonumber \\
			& + \sum_{x,I<J} B_{IJ}(x) L_{IJ}(x) + \sum_x V(n(x))
\end{align}
where the various terms correspond to, respectively, the kinetic energy of $SO(5)$ rotors ($\sim \, L^2$), the coupling between rotors on different sites ($\sim \, n^2$, $<..>$ denotes sum over nearest neighbors), coupling between an external field and the momenta of the rotors ($\sim \, B L$) and a symmetry breaking term ($V(n)$) which breaks the $SO(5)$ symmetry down to $SO(3) \times U(1)$.



\section{Cartan Decomposition}\label{sec:cartan_gravity}

We now come to the gravity side of the picture. Our ingredients are a four-dimensional manifold $\mc{M}^4$ on which we have a $ SO(4,1) $ or $ SO(3,2) $ connection $ A_{\mu}^I $, depending on whether $ \Lambda > 0 $ and $ \Lambda < 0 $ respectively. There is no metric structure on this manifold to begin with. This connection can then be decomposed into two parts \cite{Wise2009MacDowellMansouri}:
\begin{equation}\label{eqn:full_connection}
	A = \omega + \frac{\epsilon}{l} e
\end{equation}
where $ \omega_\mu^I $ is identified with an $ \mf{so}(3,1) $ connection and $ e_\mu^I $ is a four-dimensional frame field. $ \epsilon = +1 $ when $ \Lambda > 0 $ and $ \epsilon = -1 $ when $ \Lambda < 0 $. The curvature $ F[A] $ of the connection can then be written as:
\begin{equation}\label{eqn:full_curv}
	F[A]^I{}_J  = \mb{d} A^I{}_J + A^I{}_K \wedge A^K{}_J
\end{equation}
where $ I,J,K = 0,1,2,3,4 $ label the elements of the $ \mf{so(4,1)} $ (resp. $ \mf{so}(3,2) $) matrices, and the spacetime indices are suppressed. $ \mb{d} $ is the exterior derivative. In terms of these indices, \ref{eqn:full_connection} can be written as:
\begin{equation}
A^a{}_b = \omega^a{}_b; \qquad A^a{}_4 = \frac{1}{l} e^a; \qquad A^4{}_a = - \frac{\epsilon}{l} e_a
\end{equation}
where $ a,b = 0,1,2,3 $. This can be more clearly seen in the explicit matrix form:
\begin{equation}\label{eqn:so5-generators-2}
		A^I{}_J = \left( \begin{array}{ccccc}
						 0 & \omega^0{}_1 & \omega^0{}_2 & \omega^0{}_3 & e^0/l \\ 
						 \omega^1{}_0 & 0 & \omega^1{}_2 & \omega^1{}_3 & e^1/l \\ 
						 \omega^2{}_0 & \omega^2{}_1 & 0 & \omega^2{}_3 & e^2/l \\ 
						 \omega^3{}_0 & \omega^3{}_1 & \omega^3{}_2 & 0 & e^3/l \\ 
			 			 \epsilon e^0/l & -\epsilon e^1/l & -\epsilon e^2/l & -\epsilon e^3/l & 0
					  \end{array} \right) 
\end{equation}
$ \omega^0{}_i $ ($ i=1,2,3 $) are the generators of boosts, $ \omega^i{}_j $ ($ i,j = 1,2,3 $; $ i \ne j $) are the generators of spatial rotations ($\omega^0{}_1 = - \omega^1{}_0$ and $\omega^1{}_2 = - \omega^2{}_1$) and $ e^a/l $ ($ a = 0,1,2,3 $) are the generators of translations. The gauge curvature can then be expanded as follows. For the $ \mf{so}(3,1) $ part:
\begin{align}\label{eqn:so31_part}
	F^a{}_b & = \mb{d} A^a{}_b + A^a{}_c \wedge A^c{}_d + A^a{}_4 \wedge A^4{}_b \nonumber \\
			& = \mb{d} \omega^a{}_b + \omega^a{}_c \wedge \omega^c{}_d - \frac{\epsilon}{l^2} e^a \wedge e_b \nonumber \\
			& = R^a{}_b - \frac{\epsilon}{l^2} e^a \wedge e_b
\end{align}
Similary the $ \mbb{R}(3,1) $ part of the curvature is given by:
\begin{align}\label{eqn:r31_part}
	F^a{}_4 & = \mb{d} A^a{}_4 + A^a{}_b \wedge A^b{}_4 \nonumber \\
			& = \frac{1}{l} \left(\mb{d} e^a + \omega^a{}_b \wedge e^b \right) \nonumber \\
			& = \frac{1}{l} \mb{D}_\omega e^a
\end{align}
where $ \mb{D}_\omega $ is the antisymmetrized covariant derivative operator w.r.t the connection $ \omega $. Finally we see that the various components of the gauge field strength can be written as:
\begin{subequations}
	\label{eqn:cartan_decomp}
	\begin{align}
		F^a{}_b & = R^a{}_b - \frac{\epsilon}{l^2} e^a \wedge e_b 	\label{eqn:cartan_decomp_a}\\ 
		F^a{}_4 & = \frac{1}{l} \mb{D}_\omega e						\label{eqn:cartan_decomp_b}
	\end{align}
\end{subequations}
where $ R $ is the curvature of a $ \mf{so}(3,1) $ connection $ \omega $ and $ e $ is a $ \mf{so}(3,1) $ valued one-form.

\section{BF Theory}\label{sec:bftheory}

The action for a topological theory on a manifold $\mathcal{M}$ with
local gauge group $G$ is given by:
\begin{equation}
	\label{eqn:bf_action}
	S_{BF}=\int \left( B_{IJ} \wedge F^{IJ} \right)
\end{equation}
where $B$ and $F$ are a $(n-2)$-form and a $2$-form respectively on $\mathcal{M}$ and which takes values in the Lie-algebra $\mf{g}$ of $G$. $F$ is the field strength for a connection $A$. The configuration variables are the gauge connection $A$ and the two-form field $B$ and the action is invariant under $SO(5)$ transformations of the gauge field. Varying the action w.r.t. these variables we find the two equations of motion \cite{Freidel2005Quantum}:
\begin{subequations}
	\label{eqn:bf_eom}
	\begin{align}
		\frac{\delta S_{BF}}{\delta B} & = 0 \Rightarrow F_{IJ} = 0 \\ 
		\frac{\delta S_{BF}}{\delta A} & = 0 \Rightarrow \mb{D}_A B^{IJ} = 0
	\end{align}
\end{subequations}
where $F_{IJ} = \mb{d} A_{IJ} + A_I{}^K \wedge A_{KJ}$ is the curvature tensor and $\mb{D}_A$ is the covariant derivative w.r.t. the gauge connection $A$. We have made use of the fact that:
\beq
	\label{eqn:field_variation}
	\delta F_{IJ} = \mb{d} (\delta A_{IJ}) + A_I{}^K \wedge \delta A_{KJ} = \mb{D}_A (\delta A_{IJ})
\eeq
followed by a partial integration in order to obtain \eqref{eqn:bf_eom}.

Since the field strength is identically zero everywhere, in the present form, this action describes a system with no local degrees of freedom. The value of $S_{BF}$ when evaluated on a given manifold, for any choice of $B$ and $A$, will only yield information about the topology of the manifold. Thus \eqref{eqn:bf_action} is the action for a \emph{topological field theory} or TFT and as such has no correspondence with classical general relativity. The situation changes, however, when we add a term to the action quadratic in the $B$ field which corresponds to the breaking of the $SO(5)$ symmetry of the theory resulting in a theory with propagating local degrees of freedom. The modified action is as follows \cite{Freidel2005Quantum}:
\beq
	\label{eqn:modified_bf_action}
	S'_{BF}=\int \left( B_{IJ} \wedge F^{IJ} - \frac{1}{2} B^{IJ} \wedge B^{KL} \epsilon_{IJKLM} v^M \right)
\eeq
where $v^M$ is a fixed $SO(5)$ vector pointing in a preferred direction. It is this choice of a preferred direction that breaks the $SO(5)$ symmetry, in much the same way as the choice of a preferred direction for spins breaks the symmetry of the Ising model and allows the ferromagnetic phase to appear from an initially disordered phase where the spins point in arbitrary directions.

The equations of motion for the modified action are:
\begin{subequations}
	\label{eqn:modified_bf_eom}
	\begin{align}
		\frac{\delta S'_{BF}}{\delta B} & = 0 \Rightarrow F_{IJ} = \frac{1}{2} \epsilon_{IJKLM} B^{KL} v^M \\ 
		\frac{\delta S'_{BF}}{\delta A} & = 0 \Rightarrow \mb{D}_A B^{IJ} = 0
	\end{align}
\end{subequations}
Now we can always choose our co-ordinates in the $SO(5)$ space such that $v^M$ has only one non-vanishing component, such that $v^M := (0,0,0,0,\alpha/2)$. Then the equation of motion for the $B$ field in \eqref{eqn:modified_bf_eom} becomes:
\begin{subequations}
	\label{eqn:f_eom}
	\begin{align}
		F_{ab} & = \frac{\alpha}{4} \epsilon_{abcd} B^{cd} 	\label{eqn:f_eom_a} \\
		F_{a4} & = 0										\label{eqn:f_eom_b} 
	\end{align}
\end{subequations}
where $a,b \in \{0,1,2,3\}$; whereas the e.o.m for the gauge connection is unchanged. The second of these equations in combination with \eqref{eqn:cartan_decomp_b} tells us that:
\begin{equation*}
	\mb{D}_\omega e = 0
\end{equation*}
\emph{i.e.}, the torsion of the gauge connection is zero. Contracting both sides of \eqref{eqn:f_eom_a} with $\epsilon^{ef}{}_{ab}$ we obtain:
\begin{equation}\label{eqn:b_eom}
	B^{ef} = \frac{1}{\alpha} \star F^{ef}
\end{equation}
where $\star$ is the Hodge dual operator (contraction with $\epsilon_{abcd}$). Substituting the solution for $B$ \eqref{eqn:b_eom} into the modified action and using the fact that $F_{a4} = 0$ \eqref{eqn:f_eom_b}, we find:
\begin{align}
	\label{eqn:modified_bf_action_2}
	S'_{BF} & =\int \left( - \frac{1}{\alpha}\star F_{ab} \wedge F^{ab} - \frac{1}{\alpha} \star F^{ab} \wedge F_{ab} \right) \nonumber \\
	& = -\frac{1}{2\alpha} \int F^{ab} \wedge \star F_{ab} \nonumber \\
	& = -\frac{1}{2\alpha} \int \left(R^{ab} - \frac{\epsilon}{l^2} e^a \wedge e^b\right) \wedge \star \left( R_{ab} - \frac{\epsilon}{l^2} e_a \wedge e_b \right) \nonumber \\
\end{align}
where in the third line we have utilized the identity \eqref{eqn:cartan_decomp_a}. Finally we have:
\begin{align}
	S'_{BF} =  -\frac{1}{2\alpha} \int \bigg[ R^{ab} \wedge \star R_{ab} & + \frac{\epsilon^2}{l^4} e^a \wedge e^b \wedge \star (e_a \wedge e_b) \nonumber \\ 
	& - \frac{2\epsilon}{l^2} R^{ab} \wedge \star (e_a \wedge e_b) \bigg]
\end{align}
The last two terms give us the Palatini action\footnote{The connection formalism and the first order Palatini action for General Relativity are reviewed in a forthcoming review article on LQG \cite{Vaid2013Loop}.} for general relativity with a cosmological constant, while the first term is a topological term whose variation vanishes due to the Bianchi identity.

\section{Physical Interpretation}\label{sec:interpretation}

It is straightforward to see the correspondence between the operators for charge, rotations and translations (acting on the electron wavefunction which) form the components of the $SO(5)$ connection \eqref{eqn:so5-generators-1} and the operators defined in the spacetime connection given in \eqref{eqn:so5-generators-2}. First let us write down the form of the $5 \times 5$ matrix generators of the Lie algebras of $\mf{so}(4,1)$, $ \mf{iso}(3,1)$ and $ \mf{so}(3,2) $ in the following suggestive form \cite[p.~10]{Wise2009MacDowellMansouri}:
\begin{equation}\label{eqn:lie-algebra-generators}
	\left(
		\begin{array}{ccccc}
			0 	& b^1 	& b^2 	& b^3 	& p^0/l \\
			b^1 & 0 	& j^3	& -j^2	& p^1/l \\
			b^2 & -j^3 	& 0		& j^1	& p^2/l \\
			b^3 & j^2 	& -j^1	& 0		& p^3/l \\
			\epsilon p^0/l & -\epsilon p^1/l 	& -\epsilon p^2/l	& -\epsilon p^3/l	& 0 \\
		\end{array}
	\right) = j^i J_i + b^i B_i + \frac{1}{l}p^a P_a
\end{equation}
where $J_i$ are the generators of rotations, $B_i$ generate boosts and $P_a = (P_0, P_i)$ generate translations. The value of the factor $\epsilon$ determines the type of the algebra. If $\epsilon$ is $-1$ , $0$ or $1$, the Lie-algebra the above matrix describes is $\mf{so}(4,1)$, $ \mf{iso}(3,1)$ or $ \mf{so}(3,2) $ respectively.

Table~\ref{tbl:correspondence} illustrates this correspondence.

\begin{table}[h]
	\centering
	\begin{tabular}{|c|p{5em}|p{5em}|}
		\hline
			{} & $L_{IJ}$ & $A_{IJ}$ \\
		\hline
			 Rotations   &  \[ \barr{c} S_x \\ S_y \\ S_z \earr \] & \[ \barr{c} -\omega^3{}_2 \\ \;\omega^3{}_1 \\ -\omega^2{}_1 \earr \]  \\
		\hline
			Boosts		 & \beq \barr{c} \pi^\dagger_x + \pi_x \\ \pi^\dagger_y + \pi_y \\ \pi^\dagger_z + \pi_z \earr \nonumber \eeq & \beq \barr{c} \omega^1{}_0 \\ \omega^2{}_0 \\ \omega^3{}_0 \earr \nonumber \eeq \\
		\hline
			Translations & \beq \barr{c} i (\pi^\dagger_x - \pi_x) \\ i (\pi^\dagger_y - \pi_y) \\ i (\pi^\dagger_z - \pi_z) \earr \nonumber \eeq & \beq \barr{c} \epsilon e^1/l \\ \epsilon e^2/l \\ \epsilon e^3/l \earr \nonumber \eeq \\
		\hline
			Charge & $ Q $ & $ \epsilon e^0/l $ \\
		\hline
	\end{tabular}
	\caption{Physical correspondence between operators in the condensed matter system and in the gravitational theory.}
	\label{tbl:correspondence}
\end{table}

\section{Discussion: Phases of Spacetime}

On the condensed matter side, it is understood that the $SO(5)$ formalism for High-T\subsc{C} superconductivity and anti-ferromagnetism is only an approximation (or effective field theory) \cite{Burgess1996On-the-SO5-Effective,Burgess1997SO5-Invariance} that arises in the long-wavelength low-energy limit of the physics of some underlying fundamental dynamics. In \cite{Rabello1997Microscopic,Demler2004SO5-theory} several examples of microscopic Hamiltonians are given who long-wavelength theory explicitly exhibit the $SO(5)$ symmetry. A recurring example of an exact microscopic Hamiltonian in the case of High-T\subsc{C} SC/AF is that of the tight-binding Hubbard model. In \cite{Vaid2012Quantum,Vaid2013Non-abelian} we pointed out that the behavior of black hole entropy in LQG suggests a connection between the physics of a black hole horizon and that of the quantum hall effect. There we suggested the Hubbard model as a candidate microscopic Hamiltonian for describing the physics of a black hole horizon. The present work provides support for this proposal. This addresses the question of the microscopic origin of the effective $SO(5)$ theory in the gravitational context.

Knowledge of the detailed phase diagram of High-T\subsc{C} SC/AF also allows us to make concrete suggestions regarding the possible phases which spacetime geometry can manifest. The important aspect is the ability to identify the various phases - superconducting, anti-ferromagnetic, ferromagnetic, spin-bag, etc. - with the various solutions of Einstein's equations. To do so we can refer to the dictionary given in table~\ref{tbl:correspondence}.

It is important at this stage to point out a crucial difference between Zhang's $ SO(5) $ system and our model. In Zhang's model the $ SO(5) $ symmetry is a \emph{global} symmetry: the Lie-algebra generators in \eqref{eqn:so5-generators-1} do not have any spatial dependence. In our model we have gauged this symmetry and made it \emph{local}: Lie-algebra generators in \eqref{eqn:so5-generators-2} have a spacetime index (which was not shown in the text to avoid clutter). For instance the rotation generators $ \omega^i{}_j $ are more accurately written as $ \omega_\mu{}^i{}_j $ with a spacetime index $ \mu $.

In the AF phase the order parameter is given by the Neel vector $\vect{S} = (S_x, S_y, S_z)$. The dictionary table~\ref{tbl:correspondence} tells us that on the gravitational side this corresponds to the components of the $\mf{so}(5)$ connection which correspond to spatial rotations $(-\omega^3{}_2,\omega^3{}_1, -\omega^2{}_1)$ in the symmetry broken theory. Thus in order to associate a geometric configuration with an AF phase, we should look for a solution of Einstein's equations where rotations in the spatial planes are determined. An example is spacetime of a Kerr-deSitter \footnote{The theory we are considering has $\Lambda \ne 0$, thus one has to work with the deSitter/anti-deSitter generalization of the Kerr spacetime.} black hole, which describes a rotating black hole. Observers outside a Kerr-deSitter black hole will experience a spacetime with broken rotational invariance - with the rotation axis of the black hole defining a preferred direction in space - and far from the horizon, the generators of spatial rotations $(-\omega^3{}_2,\omega^3{}_1, -\omega^2{}_1)$ will reach a constant, non-zero value. Thus, the geometry experienced by observers far from the horizon of a rotating black hole can be identified with the anti-ferromagnetic phase.

For the SC phase it is, at present, not clear to us as to what geometric configuration should be identified with with it. A guess would be that the geometry near or inside the hozion of a charged - Reissner-Nordstorm - black hole can be identified with a SC phase\footnote{It is known that a scalar field living in the charged black hole background will undergo symmetry breaking leading to formation of a superconducting condensate in the near horizon region \cite{Hartnoll2008Building,Hartnoll2008Holographic,Hartnoll2010Lectures,Herzog2009Lectures}. This observation would appear to buttress our identification of a charged black hole geometry with the SC phase of a symmetry broken $ SO(5) $ theory.}. If we consider the case of black hole which is both rotating and charged - Kerr-Newman\footnote{once again with the caveat that the black hole is embedded in a bulk deSitter spacetime} - then it would appear that the AF phase can be identified with the bulk geometry far from the horizon and the SC phase with the bulk geometry in the interior of the black hole. Of course, this identification is, as yet, qualitative and requires detailed analytical investigation before it can be fully accepted. However, this does tells us the general direction one must follow for identifying phases of geometry with the phases encountered in condensed matter.


\bibliographystyle{plain}

\bibliography{phases}

\appendix

\section{Notational Conventions}\label{app:notation}

For the reader's convenience let us clarify some aspects of the notation used in this paper.

\begin{table}[h]
	\centering
	\begin{tabular}{|l|l|l|}
		\hline
		\textbf{Quantity} & \textbf{Symbols} & \textbf{Range}\\
		\hline
		5D Clifford Algebra & $I,J,K,\ldots$ & $\{0,1,2,3,4\}$ \\
		\hline
		Generators of $SO(5)$ & $IJ, JK, \ldots$ & {} \\
		\hline
		Generators of $SO(4,1)$ and $SO(3,2)$ & $a,b,c,\ldots$ & $\{0,1,2,3\}$ \\
		\hline
		Generators of $SO(3)$ and $SU(2)$ & $i,j,k,\ldots$ & $\{1,2,3\}$ \\
		\hline
		Spacetime co-ordinates & $\mu,\nu,\alpha,\beta,\ldots$ & $\{0,1,2,3\}$ \\
		\hline
	\end{tabular}
\end{table}

For the most part, spacetime indices $\mu,\nu,\ldots$ are suppressed. One-forms correspond to objects with one spacetime index: $V_\mu$. Two-forms are objects with two spacetime indices $F_{\mu\nu}$, which are antisymmetric in those indices, \emph{i.e.}, $F_{\{\mu\nu\}} = 0$, where $\{..\}$ denotes symmetrization over the enclosed indices.

The tetrad $e_{\mu}{}^I$ can be thought of as a spacetime field (labeled by $\mu$) which, at each point of our spacetime manifold, gives us a vector (labeled by $I$) - or more precisely an element of the five-dimensional Clifford algebra, which rotates under the respective gauge transformations.


The ``wedge'' product between one-forms and two-forms is defined as the completely antisymmetric outer product between two given objects. For instance, given a one-form $e_{\mu}$ and a two-form $F_{\mu\nu}$, the wedge product between the two would give a three-index object completely antisymmetric in all the indices:
\begin{equation*}
	e \wedge F \equiv \epsilon^{\mu\nu\delta} e_\mu F_{\nu\delta}
\end{equation*}
The action for BF theory written with spacetime indices shown explicitly is:
\begin{equation*}
	S_{BF} = \int d^4 x \, B^{IJ} \wedge F_{IJ} \equiv \int d^4 x \, \epsilon^{\mu\nu\alpha\beta} B_{\mu\nu}{}^{IJ} F_{\alpha \beta \, IJ}
\end{equation*}


\end{document}